\documentclass[12pt]{article}
\usepackage{amssymb,epsfig}
\begin{document}

\large

\baselineskip=36pt

\thispagestyle{empty}

 ~~

\vspace{2cm}

 {\LARGE \bf
\begin{center}
Black Hole in ADS\\
~~  \\
and Quantum Field Theory\\
\end{center} }

\bigskip

\bigskip

\begin{center}
{\Large \sc Giuseppe MAIELLA} \\
~~\\
{\sl Dept.\ of Physical Sciences \\
University of Naples ``Federico II"}
\end{center}

\bigskip

\bigskip

\begin{center}
Talk given at APC Meeting --- Paris, 10--19 December,
2007\\~~\\~~\\ \fbox{\normalsize \begin{tabular}{c} based on the
paper:\\~~\\ {\bf G. Maiella and C. Stornaiolo}, \\ ``A CFT
Description of the BTZ Black Hole:
Topology versus Geometry \\
(or Thermodynamics versus Statistical Mechanics)", \\~~\\
 {\sl
Int. J. Mod. Phys.}  {\bf 22} nr. 20 (2007) 3429--3446;~~~~
hep-th/0611194
\end{tabular}}
\end{center}

\newpage

\setcounter{page}{1}

\noindent {\Large \bf \underline{Plan of the talk}}

\bigskip

\begin{description}
\item[A)] Introduction

\bigskip

\item[B)] It is given a detailed analysis of the well-known case
ADS$_3$/CFT$_2$. The geometrical and topological properties are
presented.\\
The exact solution of unitary CFT$_2$ with central charge $c=1$
implies exact results on the Black Hole (BH) solution for the
ADS$_3$ space--time, the so-called BTZ (Ba\~nados, Teitelboim,
Zanelli, {\sl PRL} {\bf 69} (1992), 1849).

\bigskip

\item[C)] The peculiarity of BTZ Black Hole is made explicit by
showing the topological nature of $T_H$ and its relation with the
boundary (quantum) energy $E_C =c/6$.

\bigskip

\item[D)] As possible avenue for the extension of $D=2$ results to
any $D$ we analyze the moving mirror analogy of Black Hole and the
Verlinde proposal to substitute the quantum anomaly $c/6$ with the
Casimir energy $E_C$.\\
Both avenues seem to suggest that $S_H = S_{\mbox{\small ent}}$
for any $D \geq 2$. This aspect is under study by G. Maiella and
C. Stornaiolo.

\end{description}

\newpage

~~\\

\bigskip

\noindent {\Large \bf A) ~~Introduction}

\bigskip

\bigskip

I will present a short review of recent achievements in
understanding the origin of the statistical properties versus the
thermodynamical ones for Anti De Sitter space--time, ADS$_{n+1}$
from the point of view of the ``dual" field theory defined on the
boundary of ADS$_{{n+1}}$ conformal invariant, CFT$_n$.

 We try to
convince the audience that the symmetries of Quantum Field Theory
at the critical points are, on one side, at the basis of the
thermodynamical properties of Black Hole solution, i.e.\ the
Hawking temperature $T_H$ and the entropy $S_H$; on the other side
the CFT$_n$ in Euclidean time determines the statistical
properties of BH.

\newpage

\noindent {\Large \bf  B) ~~Geometry of the ADS$_3$ space--time}

\bigskip

The ADS$_3$ / CFT$_2$ case can be considered as a ``toy model",
where exact results are known.

For our purpose we consider the metric of ADS$_3$ in polar
``coordinate"

\[
ds^2 = \left[   \left( \frac{r}{l} \right)^2 + 1 \right] dt^2 +
\left[   \left( \frac{r}{l} \right)^2 + 1 \right]^{-1} dr^2 + (r d
\varphi)^2 \]

\noindent where $\Lambda = -1/l^2$.

It is easy to show that the boundary $(r \rightarrow \infty)$ is
conformal; then the metric is invariant under the diffeomorphism
which preserves the boundary conditions and their symmetries
(i.e.\ the invariance group $SO (2,2)$, or better the covering
group $SL (2, C)$). Their infinitesimal generators define the
classical Virasoro algebra.

\newpage

Therefore it is useful to write the action in terms of a
``would-be" gauge field
${\displaystyle A^{(\pm)} = \!\!\!\!\!\!\!\begin{array}[t]{c}\omega  \\
\downarrow \\ \mbox{\small spin conn.} \end{array}
\!\!\!\!\!\!\!\pm \frac{1}{e}
\begin{array}[t]{l}e \\ \downarrow \\ \mbox{\small vierbein}
\end{array}}$ so to give:

\bigskip

\[I_{CS} = \frac{k}{4  \pi} \int \, \mbox{Tr}\, \left\{ A \wedge
dA + \frac{2}{3} A \wedge A \wedge A \right\} \]

\bigskip

\noindent $I_{CS} \equiv$ Chern--Simon action; ~~~$F = dA + A
\wedge \underline{A} =0$ is a constraint and  $k$ is the ``central
charge" of Kac--Moody algebra (I won't discuss this aspect here).

\newpage

\noindent {\Large \bf BTZ Black Hole}

\bigskip

A {\bf rotating} Black Hole solution is described by
metric (with Lorentz signature)

\begin{eqnarray*}
ds^2 & = & - \left[ -M_L + \frac{r^2}{l^2} + \frac{J_L^2}{4 r^2}
\right] dt^2 \\~~\\
& & + \left[ -M_L + \frac{r^2}{l^2} + \frac{J_L^2}{4 r^2} \right]
^{-1} dr^2 + r^2 \left[ d \Phi - \frac{J_L}{2 \pi} dt \right]^2
\end{eqnarray*}

\noindent where $J_L^2$ is the angular momentum.

\bigskip

\noindent {\bf Notice} that $l$, the curvature radius for
ADS$_3$: $l$ sets the {\bf length scale} for the problem.

\noindent As usual, we impose that the lapse function

\[ N = - M_L + \frac{r^2}{l^2} + \frac{J_L^2}{4r^2} \]

\noindent is zero. \\
Then we get the value of the horizons radii $r_{\pm}$ as

\[ r_{\pm}^2 = \frac{M_L l^2}{r} \left[ 1 + \sqrt{1 - \frac{J_L^2}{M_L^2 \cdot
l^2}} \right] \]

\noindent from which the invariants are

\begin{eqnarray*}
& & M_L^2 = \left( \frac{r^2_+ + r_-^2}{l^2} \right)^2
~~~~~~~~M_L\,\, \mbox{adimensional ``mass" of BH} \\
& & ~~\\
 & & J_L^2 = 4 \left( \frac{r_+ r_-}{l^2} \right)^2
~~~~~~~~~~~~J_L\,\, \mbox{angular momentum}\\~~\\
 & & M_L^2 +
J_L^2 = \frac{(r_+ + r_-)^4}{l^4}
\end{eqnarray*}

\noindent  For Euclidean space--time one gets a similar structure
with the substitution

\[ J_L^2 \longleftrightarrow - J^2 \]

~~\\
The dilatation symmetry generators $L_0$, $\bar{L}_0$ are deeply
related to the invariants of BTZ Black Hole; in fact

\[ M_L = L_0 + \bar{L}_0 ~~~~~~~~~J_L = L_0 - \bar{L}_0\]

~~\\
Geometrically this identification follows from the form of the
two Casimir invariants of $SO (2,2)$, explicitly written in
function of the six Killing vectors.

We can see that

\bigskip

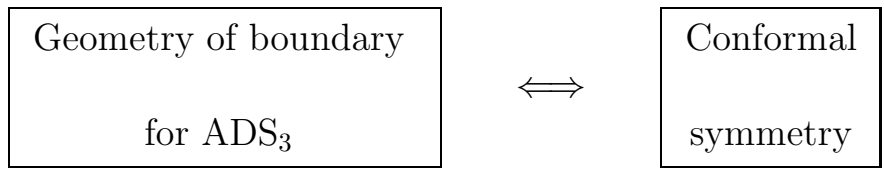
\begin{center} \fbox{\begin{tabular}{c} Geometry of boundary
\\~~\\
for ADS$_3$ \end{tabular} } ~~~~~$\Longleftrightarrow$
~~~~~\fbox{\begin{tabular}{c} Conformal \\ ~~   \\ symmetry
\end{tabular}}
\end{center}

\newpage

~~

\noindent {\Large \bf C) ~~ CFT$_2$ and Quantum effects}

\bigskip

Brown--Henneaux long ago derived in a paper [{\sl Comm. Math.
Phys.} (1986)] a striking result for ADS$_3$ space--time

~~

\centerline{\fbox{$\displaystyle c = \frac{3}{2} \, \frac{l}{G_3}
$}}

~~\\~~\\

-- $G_3$ is the Newton constant for 3D\\~~\\

-- $c$ is the central charge of the ``quantum" Virasoro algebra
for CFT$_2$ (introduced in old time $\simeq 1969$ by Virasoro)

\[ [L_n, \, L_m] = (n-m) L_{n+m} + c [n (n^2 -1)] \delta_{n+m;0} \]

\newpage

Simple derivation (Euclidean signature)

~~\\
 \noindent a) Modify $T_{\mu \nu}$ by a boundary term

\[ \hat{T}_{\mu \nu}^E = T_{\mu \nu} - \frac{1}{8 \pi G_3} \frac{1}{l}
\gamma_{\mu \nu} \]

\noindent $ \gamma_{\mu \nu} =$ boundary metric

~~\\
The asymptotic ADS$_3$ metric is given by

\[ ds^2 ~~= \!\!\!\!\!\!\!\!\!\!\raisebox{-14pt}{$\scriptscriptstyle
r^2 \rightarrow \infty$} \,\,\,\, \left( \frac{l}{r} \right)^2
(dr)^2 + \left( \frac{r}{l} \right)^2 [ (dx_1)^2 + (dx_2)^2] \]

~~\\
In complex coordinate $z = x_1 +ix_2$, $\bar{z} = x_1 - ix_2$ we
get the two sectors: left (right) correspond to analytical
(antianalytical) $z$ $(\bar{z})$ transformations; then the
diffeomorphisms are the transformations

\[ z \rightarrow z - f (z) \, ;~~~~ \bar{z} \rightarrow \bar{z} -
g (\bar{z}) \]

\noindent which preserve the boundary metric.

~~\\
b)  By requiring invariance of the metric we get for $T_{zz}$
$(T_{\bar{z} \bar{z}})$

\begin{eqnarray*}
& & T_{zz} \rightarrow T_{zz} + \ldots \ldots + \frac{l^2}{2}
(\partial_z^3 f (z)) dz^2 \\
& & T_{\bar{z} \bar{z}} \rightarrow T_{\bar{z} \bar{z}} + \ldots
\ldots + \frac{l^2}{2} \left( \partial^3_{\bar{z}} g (\bar{z})
\right) d\bar{z}^2
\end{eqnarray*}

~~\\
 One recognizes that the $\partial^3_z$ term generates a term
analogous to the ``central charge" term in the Virasoro algebra.
That brings to

\[ c = \frac{3}{2} \frac{l}{G_3} \]

~~\\
But $c$ is a ``quantum" boundary effect in CFT$_2$, and $c/6$
measures the boundary energy (the Casimir energy).

~~

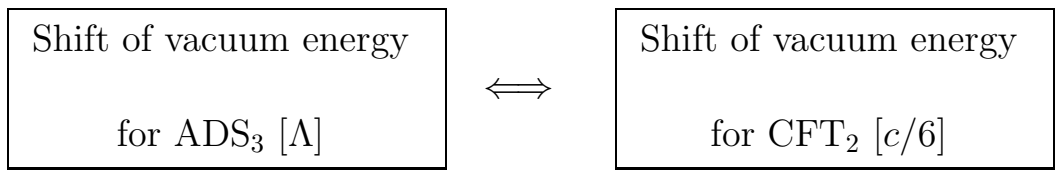
\begin{center}
\fbox{\begin{tabular}{c} Shift of vacuum energy \\~~\\ for ADS$_3$
[$\Lambda$] \end{tabular} }~~~$\Longleftrightarrow$ ~~~
\fbox{\begin{tabular}{c} Shift of vacuum energy \\ ~~\\ for
CFT$_2$ [$c/6$] \end{tabular} }
\end{center}

\newpage

\noindent {\Large \bf D) Black Hole thermodynamics for $D>2$}

\bigskip

Once identified the CFT$_2$ relevant for BTZ B.H.\ as a $c=1$
``compactified" with $R^2=1$, one can derive $T_H$ and $S_H$.

To understand how $T_H$ is related to the topology of ADS$_3$, by changing 
the coordinates in the hyperbolic 3-space $H_3$, we find

\begin{eqnarray*}
ds^2 & = & \frac{l^2}{z^2} (dx^2 + dy^2 + dz^2) = \\
& = & \frac{l^2}{\sin^2 \chi} \left( \frac{dR^2}{R^2} + d\chi^2 +
\cos^2 \chi d \vartheta^2 \right) \end{eqnarray*}

~~\\
 $\vartheta, \, \chi$ are angles, then the topology is $R_2
\times S_1$. In this coordinate system, the Euclidean time $\tau$   is periodic with period

\[ \beta_H = \frac{1}{T_H} = \frac{2 \pi r_+}{r_+^2 - r_-^2} l \]

\noindent from which for $r_- =0$ ~~(i.e. $J=0$)~~ one gets
\[S_H = \frac{1}{4} \frac{2 \pi \, r_+}{G_3} \equiv \frac{1}{4}\,
\frac{\mbox{Area}}{G_{\mbox{\small bulk}}}\] as it should be.

\newpage

One can show that $S_H$ saturates the Bekenstein bound.

In CFT$_2$ all the previous periodicity is a generator of modular
invariance $$[\tau \rightarrow \tau +1 \mathrm{\ \ transf}.]$$

\bigskip
Remind: $${\displaystyle \tau \rightarrow \frac{1}{\tau}, \, \tau
\rightarrow \tau +1}$$ generate the modular group $SL (2, Z)$,
i.e.\ the seed of all the ``dualities".

\newpage

\noindent {\Large \bf E) ~~ Thermodynamics of the black holes for
$D
>2$}

\bigskip

\bigskip

\noindent {\bf 1.  ~~~The moving mirror case}

\bigskip

I present an analysis of the moving mirror case (see C. Holzhey,
F. Larsen, Frank Wilczek {\sl Nucl. Phys. B} {\bf 424} (1994)
443-467; e-Print: hep-th/9403108, HLW from now on).

\noindent Indeed the moving mirror has been used to simulate the
quantum behavior of fields near the horizon of the black holes.

\noindent Even in four dimensions they can be described by a two
dimensional quantum field theory.

For that we exploit the conformal properties of the system.

Let's consider a  moving mirror which obeys the equation of motion
$x=f(t)$, $|\dot{f}|<1$ and   $f(t)=0$ for $t<0$.

We introduce also a massless  scalar field described by the
equation
\[
\square \phi=\frac{\partial^{2}\Phi}{\partial u \partial v}=0
\,\,\,\ \mbox{where} \,\,\,  u=x- t \, \, \, \mbox{and}\,\,\,
v=x+t \]

\noindent In a 2D Euclidean space the field equation is given by

\[
\frac{\partial^{2}\Phi}{\partial z \partial
\bar{z}}=0 \,\,\,\mbox{where} \,\,z=x+it
 \,\,\,\mbox{and}\,\,\,
\bar{z}=x-it
\]

\noindent and at the boundary the reflection condition is

\[
\phi(t,f(t))=0 \,\,\,\mbox{(the reflection condition)}
\]

\noindent To describe the effects of a moving mirror on a scalar
field, I consider its interaction with a detector through the
Lagrangian

\[
{\cal{L}}_{int.}= g k (\tau) \Phi[x(\tau)]
\]

\noindent where $g$ is a weak coupling constant and $k$  is the
monopole momentum of the detector.

The response function can be evaluated to be

\[
{\cal{F}}(E)=\int_{-\infty}^{+\infty}d\tau
\int_{-\infty}^{+\infty}d\tau' e^{-iE(\tau-\tau')}
D^{+}x(\tau)x(\tau')
\]

\noindent which gives the counting of the number of particles
found by the detector. The explicit evaluation by means of the
Wightman function $D^{+}$ (or $G^{+}$ for a massive field), gives
zero for a free field $\Phi$.

The trajectory of an accelerated mirror modifies the form of the
field in such a way that $ {\cal{F}}(E)$ is no longer equal to zero,
for the ``reflected'' field.

We choose the trajectory  $z(t)\to -t-Ae^{-2 \kappa t}+B$ as $t\to
+\infty$ to simulate the behavior in the surroundings of a black
hole. The mirror reflects only the null rays with trajectory
$v<B$. $B$ \textbf{acts like a horizon}, as all the rays with $v>B$ are not
reflected.  

For a detector moving with constant velocity $w$, it results the
creation of a thermal bath of particles with

\[
k_{B}T=\frac{\kappa }{2\pi}\left[\frac{1-w}{1+w}\right]^{1/2}
\]

One can use this model to make clear aspects of the so called
geometric entropy properties of a black hole [see HLW for
details].

\newpage

\noindent We notice that those results do not depend on $D$.

\noindent Indeed we are considering a one dimensional mirror, its
properties are described in a two dimensional Minkowski spacetime.

The conformal properties of a massless scalar field in 2D become
relevant in our case.

The trajectory
\[ f(z)=D_{1} + D_{2}e^{z/4M} \]
is equivalent to the previous
description.

This function is analogous to the coordinate transformation
assigned in the Schwarzschild geometry, where the Hawking
temperature is $$T_{H}=1/8\pi M$$, for a four dimensional black hole
with mass $M$.

\newpage

Furthermore one can study the transformation properties of a
hollow black hole described by the metrics

\[
ds^{2}=\left \{
\begin{array}{ll}
dr^{2} -d\tau^{2}-r^{2}d\Omega^{2} & \mbox{if}\,\, \tau+r \leq V_{s}\\
& \\
\lambda^{2}dt^{2} -\lambda^{-2}dr^{2}-r^{2}d\Omega^{2} & \mbox{if}\,\, \tau+r \geq V_{s}
\end{array} \right.
\]

\noindent The first metric describes the interior of the black
hole (with null coordinates $U=\tau-r$ and $V=\tau+r$).

\noindent The second metric describes the exterior region (with
null coordinates $u=\tau-r_{*}$ and $v=\tau+r_{*}$, where $r_{*}$
is defined by the equation $dr_{*}/dr = 1/\lambda^{2}$).

In terms of these null coordinates we have

\[
ds^{2}=\left\{
\begin{array}{ll}
dUdV-r^{2}d\Omega^{2} & \mbox{if}\,\, V \leq V_{s}\\
& \\
\lambda^{2}dudv-r^{2}d\Omega^{2} & \mbox{if}\,\, v \geq V_{s}
\end{array} \right.
\]

\noindent It is easy to show that one can put $V(v)=v$. One can
find $U(u)$ by demanding that along the wordline $v_{s}$ of the
shell $r$ should agree in both systems.

The result is $U=c_{1}+ c_{2}e^{-\kappa u}$, where $\kappa=1/4 M$
is the surface gravity.

\newpage

\noindent {\bf 2. ~~~ Cardy formula and Casimir energy}

\bigskip

Cardy ({\sl Nucl. Phys. B}  1986) has proved that for CFT$_2$ modular
invariance implies that the density of state is given by

\[ U = 2 \pi \sqrt{\frac{c}{6} \left(L_0 - \frac{c}{12} \right)} \]

\noindent where $c/12$ is the ``vacuum" energy for $R (L)$ sector.

$U$ {\bf satisfies} the Bekenstein bound but at the same time it
is related to the Casimir energy $E_C$ {\bf for any} $D$ as

\[ U = 2 \pi R \sqrt{E_C (2E -E_C)} \]

\[E_C = \frac{c}{12} \]

\noindent Moreover the boundary quantum term $c/12$ does not
depend on external temperature $T$.

\newpage

\noindent {\bf 3.~~~Comments on entanglement entropy}

\bigskip

For $D=2$ it has been proved (see HLW) that
\[S_{ent}= \frac{c}{6} \ln \frac{L}{a} \]
where $L$ and $a$ are respectively the infrared and ultraviolet cutoffs. Then it is easy to see that $\mathbf{S_{ent} = S_H = S_{Bek}}$, being the
bound saturated both by $S_H$ and $S_{ent}$.

On the other hand $S_{ent}$ satisfies, on quite general ground,
the area law, i.e.
\[ S_{ent} = \frac{1}{4} \frac{2 \pi r_+}{G_3}
\]

 \bigskip

 \bigskip

{\bf Question}: it is possible to extend those esults for $D
>2$?

\bigskip

\begin{center}
{\bf \sl ``There are good indications that say yes"}
\end{center}

\bigskip

\centerline{ \small work in progress by G.M. and C.S.}

\newpage

\noindent {\Large \bf Unitary representations of CFT$_2$}

\bigskip

\noindent Ginsparg: hep/th/9108028 \hfill (long)

\bigskip

\bigskip

\noindent G. Maiella, C. Stornaiolo, {\sl Int. Journ.\ of Mod.
Phys. A} {\bf 22} (2007), 3429.

\newpage

\noindent {\Large \bf Appendix}

\bigskip

Comments on the formula's (see pag. 5) from the {\bf point of
view} of CFT.

$(M,J)$ are given by eigenvalues of dilatation operator $\Delta,
\bar{\Delta}$.

For $c=1$ theory compactified $\hat{p},\, \hat{w}$ are dual
variables.

So if ${\displaystyle \hat{p} = \frac{r_+}{l}}$
~~~~${\displaystyle \hat{w} = \frac{r_-}{l}}$, one gets
\begin{eqnarray*}
& & M = \Delta + \bar{\Delta} = \frac{r_+^2 + r_-^2}{l^2} \\
& & \frac{J}{l} = \Delta - \bar{\Delta} = 2 \hat{p} \hat{w} = 2
\frac{r_+ r_-}{l^2}
\end{eqnarray*}

``Duality" relations:

\[
r_+ \leftrightarrow r_- \,\,\,\,\,\, \mbox{or}\,\,\,\,\,\, r_+
\leftrightarrow \frac{1}{l} \] descend from ${\displaystyle r_+
r_- = \frac{1}{2} lJ}$.

\end{document}